\documentclass[letter]{aa}       
\usepackage{graphicx}
\usepackage{amssymb}
\usepackage{amsmath}
\usepackage{longtable}
\usepackage{supertabular}
\usepackage{natbib}
\bibpunct{(}{)}{;}{a}{}{,}

\begin{document}

\title{Physical properties of the jet in \object{0836+710} revealed
by its transversal structure}

\author{M. Perucho \and A.P. Lobanov }

\authorrunning{Perucho \& Lobanov}

\titlerunning{Physical properties of the jet in 0836+710}

\institute{Max-Planck-Institut f\"ur Radioastronomie, Auf dem
H\"ugel 69, 53121 Bonn, Germany}

\offprints{M. Perucho, \email{perucho@mpifr-bonn.mpg.de}}

\date{Received <date> / Accepted <date>}

\abstract{
{Studying the internal structure of extragalactic jets is crucial
for understanding their physics. The Japanese-led space VLBI
project VSOP has presented an opportunity for such studies, by
reaching baseline lengths of up to 36,000\,km and resolving
structures down to an angular size of $\approx 0.3$ mas at 5 GHz.}
{VSOP observations of the jet in 0836+710 at 1.6 and 5 GHz have
enabled tracing of the radial structure of the flow on scales from
2 mas to 200 mas along the jet and determination of the
wavelengths of individual oscillatory modes responsible for the
formation of the structure observed.}
{We
apply linear stability analysis to identify the oscillatory modes with
modes of Kelvin-Helmholtz instability that match the wavelengths of
the structures observed.}
{We find that the jet structure in 0836+710 can be reproduced by
the helical surface mode and a combination of the helical and
elliptic body modes of Kelvin-Helmholtz instability. Our results
indicate that the jet is substantially stratified and different
modes of the instability grow inside the jet at different
distances to the jet axis. The helical surface mode can be driven
externally, and we discuss the implications of the driving
frequency on the physics of the active nucleus in
\object{0836+710}.} {}

}

\maketitle

\section{Introduction}

Resolving the transversal (or radial) structure in extragalactic
jets is a crucial step in our understanding of the physics of
these objects. On parsec scales, this has become feasible only
recently, using space VLBI\footnote{Very Long Baseline
Interferometry} observations with VSOP\footnote{VLBI Space
Observatory Programme} \citep{lz01}. These observations revealed
the presence of a double helical structure inside the jet of
\object{3C~273}, which can be attributed to a combination of
helical and elliptic modes of Kelvin-Helmholtz (KH) instability.
Numerical simulations further support this interpretation
\citep{pe06}. In this letter, we expand the scope of this
investigation and use VSOP observations of the radio jet in the
quasar S5\,0836+710 at 1.6 and 5\,GHz to estimate basic physical
properties of the relativistic flow, focusing specifically on the
radial profiles of its velocity and density.

The luminous quasar S5\,0836+710 at a redshift $z=2.16$ hosts a
powerful radio jet extending up to kiloparsec scales \citep{hu92}.
VLBI monitoring of the source \citep{ot98} has enabled estimates
of the bulk Lorentz factor $\gamma_\mathrm{j}=12$ and the viewing
angle $\theta_\mathrm{j}=3^\circ$ of the flow. Presence of
instability developing in the jet is suggested by the kink
structures observed on milliarcsecond scales with ground VLBI
\citep{kr90}.

In the VSOP image of 0836+710 at $5\,\rm{GHz}$, oscillations of the
jet ridge line are observed \citep{lo98}, with wavelengths of
$7.7\,\rm{mas}$ and $4.6\,\rm{mas}$.  These structures have been
identified with the helical surface and elliptic surface modes of KH
instability, yielding estimates of basic physical properties of the
flow: the Lorentz factor $\gamma_\mathrm{j}\sim11$, Mach number
$M_\mathrm{j}\sim6$ and jet/ambient medium density ratio
$\eta(\rho_\mathrm{j}/\rho_\mathrm{a})=0.04$. High dynamic range
VLBA\footnote{Very Long Baseline Array of National Radio Astronomy
Observatory, USA} observations of 0836+710 at $1.6\,\rm{GHz}$ show the
presence of an oscillation at a wavelength as long as
$\sim100\,\rm{mas}$ \citep{lo06}. This wavelength cannot be reconciled
with the jet parameters determined from the two shorter wavelength
oscillations, indicating that the flow may have a complex, stratified
transversal structure in which emission at lower frequencies
originates from outer layers of the flow.

Taking into account that the parameters given in \cite{lo98} were
obtained under the assumption of a vortex sheet transversal structure
of the jet, we explore the possibility that the jet is sheared, with
the long wavelength corresponding to a surface mode growing in the
outer layers, whereas the shorter wavelengths are identified with body
modes developing in the inner radii of the jet. In our approach, we
assume that the magnetic field is not dynamically important and that
the structure of the jet is due to KH instability. Under
these assumptions, the jet structure and evolution can be described within
the framework of linear stability theory for cylindric outflows.

In Section~2, we describe the method used to solve the linear
stability problem for cylindric relativistic flows and provide the
respective solutions for the set of parameters given in
\cite{lo98}. This solutions are compared in Section~3 with the
wavelengths observed in the jet of 0836+710. Section 4 summarizes
the main results of this work and puts them in the broader context
of the physics of relativistic outflows.

\section{Linear analysis}

\begin{figure*}[!t]
\centerline{
\includegraphics[width=0.4\textwidth,angle=0,clip=true]{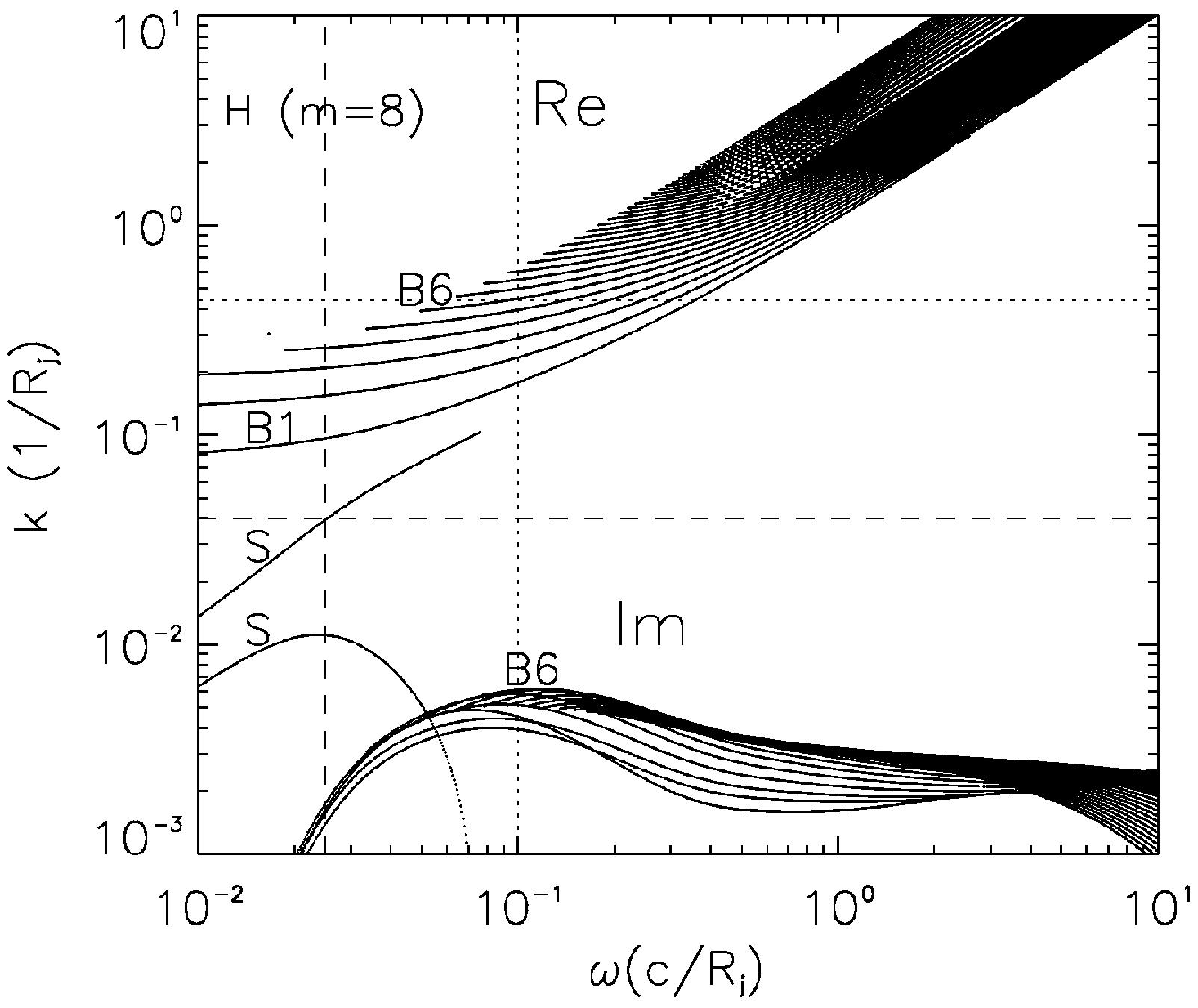}
\includegraphics[width=0.4\textwidth,angle=0,clip=true]{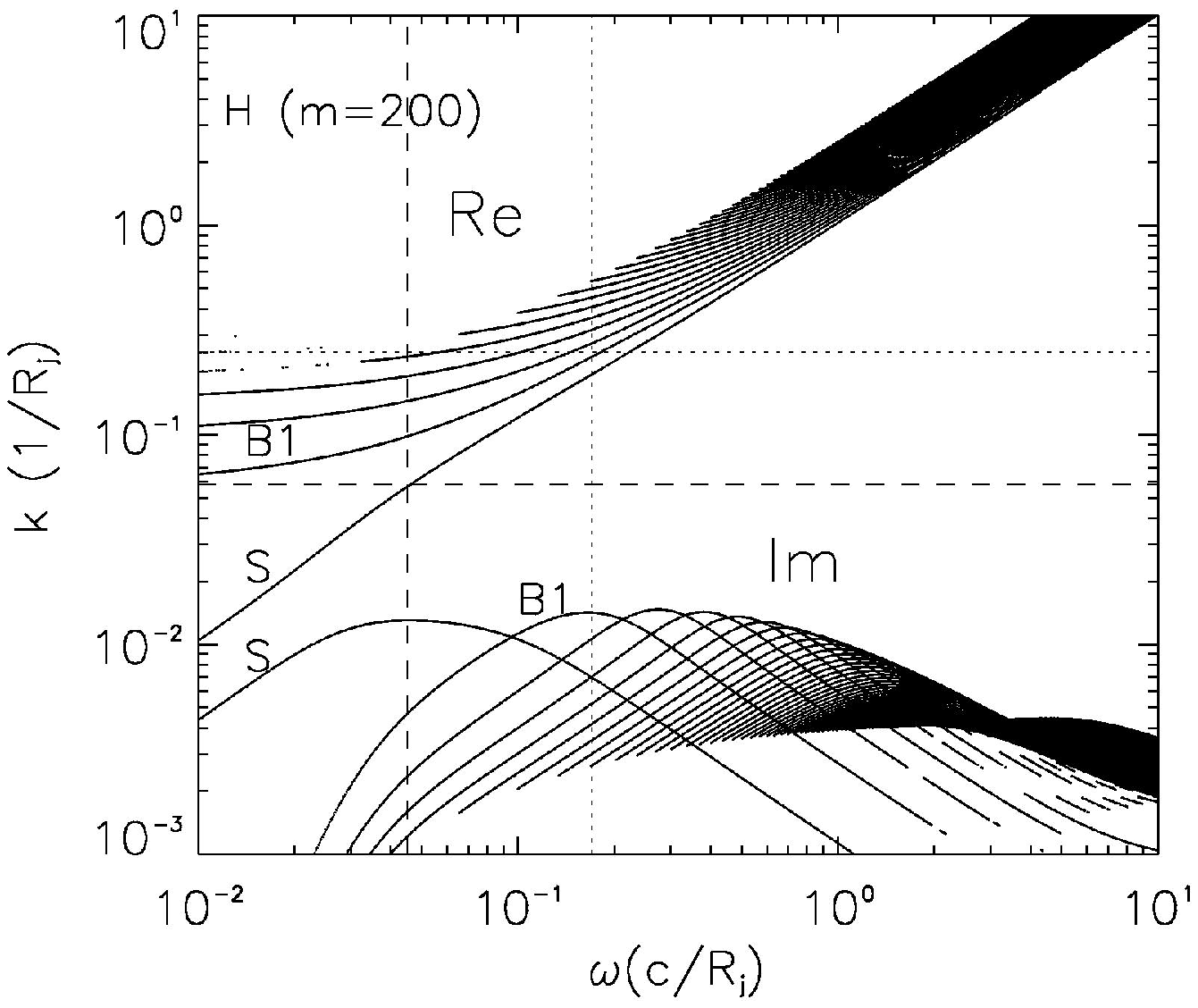}}
\caption{Real (upper curves) and imaginary (lower curves) parts of
the wavenumber versus frequency for the helical modes and the
parameters given for the jet in 0836+710. The vertical dashed
lines indicate the minimum growth lengths of the surface modes,
whereas the vertical dotted lines indicate the minimum growth
length of the first body modes for $m=200$ and the minimum growth
length of all body modes for $m=8$. The horizontal lines indicate
the wave number at which the minimum growth length occurs. $S$
stands for the surface mode, $B1$ for the first body mode, and
$B5$-$B6$ indicate the body mode that gives the minimum growth
lengths for $m=8$.} \label{fig:sols}
\end{figure*}

We describe the flow in cylindrical coordinates ($r$, $z$,
$\phi$), with $r$ and $z$ defining the radial and axial
directions, respectively. We consider a sheared transition in the
axial velocity, $v_z$ and the rest mass density, $\rho$, between
the jet and the ambient medium of the following form (see
\citealt*{pe07} for a detailed description of the linear analysis)
$a(r) = a_\infty + (a_0-a_\infty)/\cosh (r^m)$, where $a(x)$ is
the profiled quantity ($v_z$ or $\rho$) and $a_0$, $a_\infty$ are
its values at the jet symmetry plane (at $r=0$) and at $r
\rightarrow \infty$, respectively. The integer $m$ controls the
steepness of the shear layer.  In the limit $m \rightarrow
\infty$, the configuration turns into the vortex-sheet case, as
described by \cite{ha00}. The jet and the ambient medium are in
pressure equilibrium. An adiabatic perturbation of the form
$\propto g(r)\exp [i(k_z z - \omega t)]$ is introduced in the
equations of the flow. Here, $\omega$ and $k_{z}$ are the
frequency and longitudinal wavenumber of the perturbation,
respectively, and the function $g(r)$ defines the radial structure
of the perturbation. The resulting pressure perturbation $P_1$ is
then obtained from a second order ordinary differential equation
(equation 13 in \citeauthor{Bi84} \citeyear{Bi84}). This equation
can be solved by integrating from the jet axis to a point outside
the jet, where boundary conditions on the amplitude of pressure
perturbation and its first derivative are imposed. We apply the
following boundary conditions: {\em a)}~symmetry or antisymmetry
of the perturbation on the jet axis, {\em b)}~the Sommerfeld
condition requiring the solutions to approach zero at infinity,
and {\em c)}~no incoming waves at infinity.  These conditions
ensure that the solutions can be given by Bessel functions inside
the jet and Hankel functions outside. The solutions are obtained
in the spatial domain, i.e., assuming $\omega$ is real and $k$ is
complex.  In this description, $1/{\cal I}(k)$ gives the growth
length (or {\em e}-folding length), defined as the distance at
which the perturbation amplitude increases by one exponential
factor. We adopt the jet parameters derived in \cite{lo98,lo06}
and solve the pressure perturbation equation with several
different values of $m$. The resulting solutions are plotted in
Fig.~\ref{fig:sols} for the helical instability modes and for two
values of $m$: $m=8$ (left panels) and $m=200$ (right panels). The
$m=8$ case corresponds to a broad shear layer of
$\sim0.6\,R_\mathrm{j}$ in width, whereas the $m=200$ one implies
a shear layer width of $\sim0.1\,R_\mathrm{j}$, i.e., approaching
the vortex-sheet case.


The plots in Fig.~\ref{fig:sols} indicate that, in thicker shear
layers ($m=8$), the surface mode grows faster than the body modes.
The minimum growth lengths of the surface mode indicated by
vertical dashed lines in Fig.~\ref{fig:sols}) are realized at long
wavelengths (horizontal dashed lines show the wavenumber of the
mode): $\lambda^*_\mathrm{S,H,8}\sim160\,R_\mathrm{j}$. The
smallest growth lengths of all body modes is achieved by the fifth
and sixth order body modes at
$\lambda_\mathrm{B6,8}\sim14\,R_\mathrm{j}$, whereas the minimum
growth length of the first body mode (expected to be among the
most prominent in the jet) is realized at
$\lambda_\mathrm{B1,8}\sim40\,R_\mathrm{j}$.

For a thin shear layer ($m=200$), the solutions approach the
vortex-sheet approximation at the lowest frequencies (longest
wavelengths). The minimum growth lengths of low order helical body
modes are similar to that of the helical surface mode, with the
minimum realized for the first body mode at
$\lambda_\mathrm{B1,200}\sim25\,R_\mathrm{j}$. The minimum growth
length of the surface modes shifts to higher frequencies and
shorter wavelengths
($\lambda^*_\mathrm{S,H,200}\sim100\,R_\mathrm{j}$) compared to
the respective values obtained for $m=8$.  For $m<8$, the growths
of all the modes are strongly reduced as the width of the shear
layer is increased. For $m>200$ the solutions converge to the
vortex-sheet case also at the highest frequencies/shortest
wavelengths.

\begin{table*}[!t]
\caption{Intrinsic wavelengths corresponding to the observed
structures in the jet of 0836+710}
\label{tab:t1}
\begin{center}
\begin{tabular}{c|ccccc}\hline\hline
&$\lambda_\mathrm{int}$(100 mas) &$\lambda_\mathrm{int}$(100 mas) &
$\lambda_\mathrm{int}$(7.7 mas) &$\lambda_\mathrm{int}$(7.7 mas) &$\lambda_\mathrm{int}$(7.7 mas) \\
&($v_\mathrm{w}=0.6\,c$)&($v_\mathrm{w}=0.8\,c$)&($v_\mathrm{w}=0.2\,c$)&($v_\mathrm{w}=0.5\,c$)&$(v_\mathrm{w}=0.7\,c$)\\
\hline
$R_\mathrm{j,1.6}$(17 mas)&$140\,R_\mathrm{j}$&$72\,R_\mathrm{j}$& $22\,R_\mathrm{j}$&$14\,R_\mathrm{j}$ & $8\,R_\mathrm{j}$\\
$R_\mathrm{j,5}$(0.6 mas)&$4\,10^3\,R_\mathrm{j}$&$2\,10^3\,R_\mathrm{j}$&$620\,R_\mathrm{j}$&$390\,R_\mathrm{j}$&$230\,R_\mathrm{j}$\\
\hline
\end{tabular}
\end{center}
Notes: The wave speeds are given at the minimum growth length of
the relevant modes: the helical surface mode for $m=8$
($v_\mathrm{w}=0.6\,c$) and $m=200$ ($v_\mathrm{w}=0.8\,c$), the
fifth and sixth body mode for $m=8$ ($v_\mathrm{w}=0.2\,c$), and
the first body mode for $m=8$ ($v_\mathrm{w}=0.5\,c$) and $m=200$
($v_\mathrm{w}=0.7\,c$).
\end{table*}

\section{Results}

The solutions of the linear stability problem obtained in the
previous section can be compared with the wavelengths observed in
the jet in 0836+710.  This comparison requires an estimate of the
radius of the jet, in order to convert the frequencies and
wavenumbers of the solution into physical units.  The true jet
radius can be estimated from the VLBA radio maps presented in
\cite{lo98,lo06}, using the following relation $R_\mathrm{j}
(\mathrm{mas}) = 0.5 \sqrt{D_i^2-b^2}$, where $D_i$ is the
observed width of the jet and $b$ is the beam width transversally
to the direction of the jet.  We measure the diameter of the jet
at $1\%$ of the peak emission \citep{we92} at the base of the jet,
implicitly assuming that the outer parts of the shear layer emit
less than 1\% of the radio power at these frequencies. This yields
the jet radii of $R_\mathrm{j,1.6}\sim17$\,mas and
$R_\mathrm{j,5}\sim0.64$\,mas at 1.6\,GHz and 5\,GHz,
respectively. With these values, the intrinsic (rest frame)
wavelengths, $\lambda_\mathrm{int}^\mathrm{mod}$, can be obtained
and expressed in units of $R_\mathrm{j}$.  In order to compare
these wavelengths with the observations, the observed wavelengths
must be corrected for relativistic motion, projection and
cosmological effects, yielding:
$\lambda_\mathrm{int}=\lambda_\mathrm{obs}\,(1+z)\,
(1-v_\mathrm{w}/c\,\cos\theta_\mathrm{j})/ \sin\theta_\mathrm{j}$,
where $\lambda_\mathrm{obs}$ is the observed wavelength, $z$ is
the redshift, $v_\mathrm{w}$ is the wave speed and
$\theta_\mathrm{j}$ is the jet angle to the line of sight. The
wave speed $v_\mathrm{w}=\omega/k$ is obtained from the solutions
to the linear problem.  The wave speed of the helical surface mode
(most likely responsible for the longest observed wavelength of
100\,mas) ranges from $v_\mathrm{w}\sim0.6\,c$ to
$v_\mathrm{w}\sim0.8\,c$, depending on the thickness of the shear
layer.  The dominant body modes (responsible for the shorter
observed wavelengths of 7.7\,mas and 4.6\,mas) have speeds ranging
from $v_\mathrm{w}\sim0.2\,c$ to $v_\mathrm{w}\sim0.7\,c$.
Table~\ref{tab:t1} gives the intrinsic wavelengths for two main
observed wavelengths (100\,mas and 7.7\,mas), calculated with
different appropriate wave speeds in units of the jet radius. The
respective intrinsic wavelengths of the $4.6$ mas mode can be
obtained by dividing the values for the $7.7$ mas mode by a factor
of 1.7.

We note that the wavelengths obtained for the $m=200$ case cannot
be reproduced by any combination of excited KH modes at, or close
to, their resonant frequencies, although the lack of error
estimations prevents us from completely ruling out this
possibility. This fact alone points towards the jet being
transversally stratified. Further evidence for this comes from a
detailed analysis of the broad shear layer case ($m=8$).

For the $m=8$ case and for
$R_\mathrm{j,5}$, the intrinsic wavelengths for the longest structure
(100\,mas) have no counterparts in the solutions to the linear
problem, as already pointed out by \cite{lo06}. This suggests that the
radiating particles emitting at $5\,\rm{GHz}$ occupy a central spine
in a stratified jet, while the longer-wavelength structure seen at
1.6\,GHz is generated in outer layers of the flow.

The intrinsic wavelengths obtained for $R_\mathrm{j,1.6}$ are
close to the most unstable wavelengths provided by the $m=8$ shear
layer. In this case, the helical $100\,\rm{mas}$
$(140\,R_\mathrm{j})$ structure, corresponding to a wavenumber
$k\sim0.045$, can be identified with
$\lambda^*_\mathrm{S,H}\sim160\,R_\mathrm{j}$
($k^*_\mathrm{S,H}\sim0.04$).  The helical $7.7\,\rm{mas}$
$(22\,R_\mathrm{j})$ structure can be identified with the fifth or
sixth body mode excited at
$\lambda>\lambda^*_\mathrm{B6,H}\sim14\,R_\mathrm{j}$, and the
elliptic $4.6\,\rm{mas}$ $(13\,R_\mathrm{j})$ structure would
correspond to the fifth and sixth elliptic body modes growing at
$\lambda^*_\mathrm{B6,E}\sim14\,R_\mathrm{j}$, as indicated by the
linear solution (not shown). We consider this a better
explanation, and we adopt $R_\mathrm{j,1.6}$ as the generic jet
radius $R_\mathrm{j}$ in the following discussion.

One possible alternative to the stratified jet scenario is to
consider deviations of the basic jet parameters from those derived
in \cite{lo98,lo06}. To assess this alternative, we have
calculated the solutions of the stability equation for different
sets of parameters, varying the jet Lorentz factor, rest-mass
density ratio and specific internal energy. Our conclusions are
summarized below: {\em 1)}~We find that for colder jets (with a
sound speed $c_\mathrm{s,j}\sim0.054$, corresponding to $M\sim18$)
the growth lengths ($k_i^{-1}\geq10^{3}\,\rm{R_\mathrm{j}}$) of
the surface and first body modes are too large to explain the
growth of the instability in the jet. {\em 2)}~For hotter jets
($c_\mathrm{s,j}\sim0.55$, $M\sim1.8$), no body modes exist that
could reproduce the observed 7.7\,mas and 4.6\,mas wavelengths. At
the same time, the surface modes alone cannot reproduce the three
wavelengths observed.  {\em 3)}~Lighter jets
($\eta=4\cdot10^{-4}$) produce solutions similar to those arising
for colder jets, with the exception that the growth lengths are
slightly shorter in this case, but still of the order of
$500\,\rm{R_\mathrm{j}}$, i.e., even longer than the longest
observed wavelength taking the radius of the jet at
$1.6\,\rm{GHz}$. {\em 4)}~Denser jets ($\eta=0.4$) show the
surface and first body modes at the appropriate wavelengths, but
the surface mode grows at a too long growth length ($k_i^{-1}\sim
500\,\rm{R_\mathrm{j}}$).  Moreover, the observations indicate
that the surface mode must grow faster than the body modes,
contrary to the results derived for a dense jet. {\em 5)}~Slower
jets ($\gamma_\mathrm{j}=2$) show very short growth lengths for
the surface mode ($k_i^{-1}\sim17\rm{R_\mathrm{j}}$) and the
solutions yield too short wavelengths for the body modes.  {\em
6)}~Finally, much faster jets ($\gamma_\mathrm{j}=100$) produce
too long growth lengths.  This analysis indicates that the jet in
0836+710 must have a Lorentz factor close to 12 or slightly
smaller, a density ratio of $10^{-2}$--$10^{-1}$ and a sound speed
$c_\mathrm{s,j}\sim0.2$ or slightly larger --- not deviating much
from the parameters derived in \cite{lo98,lo06}.

\section{Discussion}

The description developed in this paper explains all three major
oscillations observed in the jet of 0836+710, if the flow speed
and density are both transversely stratified.  The longest
observed wavelength of 100\,mas represents the helical surface
mode, whereas the two shorter ones (7.7\,mas and 4.6\,mas) are
identified with the fifth/sixth order helical and elliptic body
modes, respectively. It is also possible that the short
wavelengths correspond to a lower order body mode (first body
mode, for example) growing not at a resonant wavelength.

It has been shown by \cite{ha94}that the helical surface mode of
KH instability can be driven by an external periodic process. In
this case, coupling this process to the helical surface mode
requires the driving frequency to be lower than the resonant
frequency. For 0836+710, the driving frequency of the helical
surface mode at its minimum growth length is
$0.025\,c/R_\mathrm{j}$ would imply a driving period of
$T_\mathrm{dr}\sim5.6\cdot10^7\,\rm{yrs}$. This is similar to the
driving period of $\sim2\cdot10^7\,\rm{yrs}$ found for
\object{3C\,449} \citep{ha94}.

The period $T_\mathrm{dr}$ can be produced by a number of physical
processes, including a binary black hole and a misaligned torus as
the most plausible \citep{app96}.  The latter can however be
discarded after \cite{lu05} have shown that the absolute magnitude
of the galaxy in the B band anticorrelates with the precession
periods due to misalignment between the black hole rotation axis
and the outer regions of the accretion disk.  The absolute
magnitude of 0836+710 in the R band is $M_R=-29.9$ \citep{ca02}.
Assuming a similar magnitude for the B-band yields precession
periods of $10^3-10^5\,\rm{yrs}$, depending on the accretion disk
and black-hole properties \citep{lu05}. This is much shorter than
$T_\mathrm{dr}$ we have derived.  Combining $T_\mathrm{dr}$ with
the estimate of the black hole mass in 0836+710
($2\cdot10^8-10^9\,\rm{M_\odot}$; \citeauthor{ta00}
\citeyear{ta00}) yields a companion mass of
$10^4-10^7\,\rm{M_\odot}$, depending on the separation between the
two black holes ($10^{17}-10^{18}\,\rm{cm}$).

Also, the kiloparsec-scale structure of 0836+710 shows a large,
decollimated feature \citep{hu92}, without any emission between it
and the $1.6\,\rm{GHz}$ jet.  It remains to be tested by new high
dynamic range observations whether the long instability mode
ultimately causes the disruption of the flow, and produces this
decollimation.

\begin{acknowledgements}
This work was supported in part by the Spanish Direcci\'on General
de Ense\~nanza Superior under grants AYA-2001-3490-C02 and
AYA2004-08067-C03-01. M.P. is supported by a postdoctoral
fellowship of the Generalitat Valenciana (\it{Beca Postdoctoral
d'Excel$\cdot$l\`encia}).
\end{acknowledgements}

\end{document}